\documentclass[12pt]{article}
\usepackage{epsfig}

\usepackage{slashed}
\newcommand{\notp}{{\slashed{p}}}
\newcommand{\re}{\mathop{\mathrm{Re}}\nolimits}
\newcommand{\im}{\mathop{\mathrm{Im}}\nolimits}

\begin{document}

\title{
\vskip-3cm{\baselineskip14pt
\centerline{\normalsize DESY 08-001\hfill ISSN 0418-9833}
\centerline{\normalsize NYU-TH/08/01/04\hfill}
\centerline{\normalsize January 2008\hfill}}
\vskip1.5cm
\bf Pole Mass, Width, and Propagators of Unstable Fermions}

\author{Bernd A. Kniehl\thanks{Electronic address:
{\tt bernd.kniehl@desy.de}.}\\
{\normalsize\it II. Institut f\"ur Theoretische Physik, Universit\"at Hamburg,}
\\
{\normalsize\it Luruper Chaussee 149, 22761 Hamburg, Germany}\\
Alberto Sirlin\thanks{Electronic address: {\tt alberto.sirlin@nyu.edu}.}\\
{\normalsize\it Department of Physics, New York University,}\\
{\normalsize\it 4 Washington Place, New York, New York 10003, USA}}

\date{}

\maketitle

\begin{abstract}
The concepts of pole mass and width are extended to unstable fermions in the
general framework of parity-nonconserving gauge theories, such as the Standard
Model.
In contrast with the conventional on-shell definitions, these concepts are
gauge independent and avoid severe unphysical singularities, properties of
great importance since most fundamental fermions in nature are unstable
particles.
General expressions for the unrenormalized and renormalized dressed
propagators of unstable fermions and their field-renormalization constants are
presented.

\medskip

\end{abstract}

\newpage

The conventional definitions of mass and width of unstable bosons are
\begin{eqnarray}
m_\mathrm{os}^2&=&m_0^2+\re A(m_\mathrm{os}^2),
\label{eq:osm}\\
m_\mathrm{os}\Gamma_\mathrm{os}&=&-\frac{\im A(m_\mathrm{os}^2)}
{1-\re A^\prime(m_\mathrm{os}^2)},
\label{eq:osw}
\end{eqnarray}
where $m_0$ is the bare mass and $A(s)$ is the self-energy in the scalar case
and the transverse self-energy in the vector boson case.
The subscript $\mathrm{os}$ means that Eqs.~(\ref{eq:osm}) and (\ref{eq:osw})
define the on-shell mass and the on-shell width, respectively.

However, it was shown in Ref.~\cite{Sirlin:1991fd} that, in the context of
gauge theories, $m_\mathrm{os}$ and $\Gamma_\mathrm{os}$ are gauge dependent
in next-to-next-to-leading order.
It has also been emphasized that Eq.~(\ref{eq:osw}) leads to serious
unphysical singularities if $A(s)$ is not analytic in the neighborhood of
$m_\mathrm{os}^2$.
This occurs, for example, when $m_\mathrm{os}$ is very close to a physical
threshold \cite{Fleischer:1980ub,Bhattacharya:1991gr,Kniehl:2000kk} or, in the
resonance region, when the unstable particle is coupled to massless quanta, as
in the cases of the $W$ boson and unstable quarks \cite{Passera:1998uj}.

In order to solve these severe difficulties, it was proposed in
Ref.~\cite{Sirlin:1991fd} to base the definitions of mass and width of
unstable boson on the complex-valued position of the propagator's pole, namely
\begin{equation}
\overline{s}=m_0^2+A(\overline{s}).
\label{eq:sbar}
\end{equation}
Writing $\overline{s}=m^2-im\Gamma$ and taking the real and imaginary parts
of Eq.~(\ref{eq:sbar}), the pole mass $m$ and the pole width $\Gamma$ of
unstable bosons are defined by the relations
\begin{eqnarray}
m^2&=&m_0^2+\re A(\overline{s}),
\label{eq:polem}\\
m\Gamma&=&-\im A(\overline{s}).
\label{eq:polew}
\end{eqnarray}

Over the last several years, a number of authors have also advocated the use
of $\overline{s}$ as the basis for the definition of mass and width
\cite{Willenbrock:1991hu}.
If one expands $A(\overline{s})$ about $m^2$ and retains only leading terms in
$\Gamma$, Eqs.~(\ref{eq:polem}) and (\ref{eq:polew}) lead back to
Eqs.~(\ref{eq:osm}) and (\ref{eq:osw}).
Thus, Eqs.~(\ref{eq:osm}) and (\ref{eq:osw}) may be regarded as the
narrow-width approximation of Eqs.~(\ref{eq:polem}) and (\ref{eq:polew}).
The great advantage of Eqs.~(\ref{eq:polem}) and (\ref{eq:polew}) is that
$\overline{s}$ is expected to be gauge independent, since it is the position
of a singularity in $S$-matrix elements.
In fact, formal proofs of the gauge independence of $\overline{s}$ [as well as
of the gauge dependence of Eq.~(\ref{eq:osm})], based on the Nielsen
identities, have been presented in the literature \cite{Gambino:1999ai}.

An expression equivalent to Eq.~(\ref{eq:polew}) is given by
\begin{eqnarray}
m\Gamma&=&-Z\im A(m^2),
\label{eq:mwz}\\
Z&=&\frac{1}{1+\im\left[A(\overline{s})-A(m^2)\right]/(m\Gamma)}.
\label{eq:z}
\end{eqnarray}
Indeed, inserting Eq.~(\ref{eq:polew}) into Eq.~(\ref{eq:z}),
Eq.~(\ref{eq:mwz}) becomes a mathematical identity.
Comparing with the conventional expression in Eq.~(\ref{eq:osw}), we note two
important changes:
$\im A(m^2)$ is evaluated at the pole mass $m^2$ rather than the on-shell mass
$m_\mathrm{os}^2$, and the derivative $-\re A^\prime(m_\mathrm{os}^2)$ in the
denominator of Eq.~(\ref{eq:osw}) is replaced by a finite difference
$\im\left[A(\overline{s})-A(m^2)\right]/(m\Gamma)$.
As explained in Ref.~\cite{Kniehl:2000kk}, this feature solves the threshold
singularities mentioned before.

Other important consequences of Eqs.~(\ref{eq:polem})--(\ref{eq:z}) are the
following:
\begin{enumerate}
\item It has been shown that the alternative definitions
$\overline{m}=(m^2+\Gamma^2)^{1/2}$, $\overline{\Gamma}=m\Gamma/\overline{m}$,
constructed from the gauge-independent parameters $m$ and $\Gamma$, can be
identified with the measured mass and width of the $Z^0$ boson
\cite{Sirlin:1991fd,Willenbrock:1991hu}.
\item Comparison of the pole and on-shell definitions of mass and width leads
to the conclusion that the gauge dependences of the latter can be numerically
very large, particularly in the case of a heavy Higgs boson
\cite{Kniehl:1998fn}.
\item It has been shown that Eqs.~(\ref{eq:mwz}) and (\ref{eq:z}) can be
obtained by imposing a suitable normalization condition on the imaginary part
of the renormalized self-energy, and that Eq.~(\ref{eq:z}) can be identified
with the field renormalization constant for unstable bosons
\cite{Kniehl:2001ch,Nekrasov:2001nh}.
\end{enumerate}

The aim of this paper is to extend the concepts of pole mass and width to 
unstable fermions in the general framework of parity-nonconserving gauge
theories, to obtain the expressions analogous to
Eqs.~(\ref{eq:polem})--(\ref{eq:z}), and to derive their unrenormalized and
renormalized dressed propagators and field-renormalization constants.
Given the fact that, with the exception of the electron, the lightest
neutrino, and the proton (or, at the elementary level, the $u$ quark), all
known fundamental fermions in nature are unstable particles, these
concepts and expressions are indeed of great significance.
In order to simplify the discussion, in the following analysis we will
disregard flavor mixing.
In the quark sector this means that we are considering a simplified theory in
which the Cabibbo-Kobayashi-Maskawa quark mixing matrix is replaced by the
unit matrix, while in the leptonic sector absence of flavor mixing naturally
occurs if the neutrino masses are neglected.

On covariance grounds, the fermion self-energy is of the form
\begin{eqnarray}
\Sigma(p)=\Sigma_+(p)a_++\Sigma_-(p)a_-,
\nonumber\\
\Sigma_\pm(p)=\notp B_\pm(p^2)+m_0A_\pm(p^2),
\end{eqnarray}
where $a_\pm=(1\pm\gamma_5)/2$ are the right/left-handed chiral projectors,
and the fermion propagator is
\begin{equation}
iS(p)=\frac{i}{\notp-m_0-\Sigma(p)}.
\label{eq:pro}
\end{equation}
Evaluating the inverse of the denominator in Eq.~(\ref{eq:pro}), one finds
\begin{eqnarray}
S(p)&=&\frac{1}{D(p^2)}\left\{\{\notp[1-B_+(p^2)]+m_0[1+A_-(p^2)]\}a_+\right.
\nonumber\\
&&{}+\!\!\left.\{\notp[1-B_-(p^2)]+m_0[1+A_+(p^2)]\}a_-\right\},\quad
\label{eq:inv}\\
D(p^2)&=&[1-B_+(p^2)][1-B_-(p^2)]\left[p^2-m_0^2f(p^2)\right],\quad
\label{eq:d}\\
f(p^2)&=&\frac{[1+A_+(p^2)][1+A_-(p^2)]}{[1-B_+(p^2)][1-B_-(p^2)]}.
\nonumber
\end{eqnarray}
The functions $A_\pm(p^2)$ and $B_\pm(p^2)$ are generally complex for
$p^2>s_\mathrm{thr}$, the threshold of virtual particles contributing to
$\Sigma(p)$.
In the case of unstable fermions, $s_\mathrm{thr}<m^2$.
It is instructive at this stage to consider the effect of parity
($\mathcal{P}$), charge conjugation ($\mathcal{C}$), and $\mathcal{CP}$
transformations.
One readily finds
\begin{eqnarray}
S(p)&\stackrel{\mathcal{P}}{\to}&\gamma^0S(p^\prime)\gamma^0,
\label{eq:p}\\
S(p)&\stackrel{\mathcal{C}}{\to}&CS^T(-p)C^{-1},
\label{eq:c}\\
S(p)&\stackrel{\mathcal{CP}}{\to}&\gamma^0CS^T(-p^\prime)C^{-1}\gamma^0,
\label{eq:cp}
\end{eqnarray}
where $p^\prime=(p^0,-\vec{p}\,)$, $C=i\gamma^2\gamma^0$, and $T$ means
{\it transpose}.
If parity is conserved, Eq.~(\ref{eq:p}) leads to $A_-(p^2)=A_+(p^2)$ and
$B_-(p^2)=B_+(p^2)$, as expected.
If the theory is invariant under charge conjugation, but not parity,
Eq.~(\ref{eq:c}) tells us that $B_-(p^2)=B_+(p^2)$ with no restrictions on
$A_\pm(p^2)$.
If the theory is invariant under the $\mathcal{CP}$ transformation, but not
under charge conjugation or parity separately, Eq.~(\ref{eq:cp}) leads to
$A_-(p^2)=A_+(p^2)$ with no restrictions on $B_\pm(p^2)$.
As expected, the latter conclusion also follows if the theory is invariant
under $\mathcal{T}$ (time reversal), while no restrictions on $B_\pm(p^2)$ or
$A_\pm(p^2)$ are derived by invoking invariance under the $\mathcal{TCP}$
transformation.

Introducing the definitions
\begin{eqnarray}
\Sigma_{1,2}(p)&=&\frac{1}{2}[\Sigma_+(p)\pm\Sigma_-(p)],
\nonumber\\
A_{1,2}(p^2)&=&\frac{1}{2}[A_+(p^2)\pm A_-(p^2)],
\end{eqnarray}
so that
\begin{eqnarray}
\Sigma_\pm(p)&=&\Sigma_1(p)\pm\Sigma_2(p),
\nonumber\\
A_\pm(p^2)&=&A_1(p^2)\pm A_2(p^2),
\end{eqnarray}
Eq.~(\ref{eq:inv}) can be written in the alternative form
\begin{eqnarray}
S(p)&=&
\frac{1}{C(p)[\notp-m_0-\Sigma_1(p)]
-\Sigma_2(p)[\Sigma_2(p)-2m_0A_2(p^2)]}
[C(p)-\Sigma_2(p)\gamma_5],
\nonumber\\
C(p)&=&\notp-\Sigma_1(p)+m_0[1+2A_1(p^2)].
\end{eqnarray}
Multiplying numerator and denominator by $C^{-1}(p)$ on the left, we obtain
the compact expression:
\begin{equation}
S(p)=\frac{1}{\notp-m_0-\Sigma_\mathrm{eff}(p)}[1-\Sigma_P(p)\gamma_5],
\label{eq:s}
\end{equation}
where $\Sigma_\mathrm{eff}(p)$ is an {\it effective} self-energy defined by
\begin{equation}
\Sigma_\mathrm{eff}(p)=\Sigma_1(p)
+\frac{\Sigma_2(p)[\Sigma_2(p)-2m_0A_2(p^2)]}{C(p)},
\label{eq:eff}
\end{equation}
and
\begin{equation}
\Sigma_P(p)=\frac{\Sigma_2(p)}{C(p)}.
\end{equation}

The position $\notp=M$ of the pole is given by
\begin{equation}
M=m_0+\Sigma_\mathrm{eff}(M).
\label{eq:M}
\end{equation}
In order to express $M$ in terms of the original self-energies, $\Sigma_1(p)$
and $\Sigma_2(p)$, and the functions $A_1(p^2)$ and $A_2(p^2)$, we note that
$M$ appears on the l.h.s.\ of Eq.~(\ref{eq:M}) and in $C(M)$ in the
denominator of the second term in $\Sigma_\mathrm{eff}(M)$ [cf.\
Eq.~(\ref{eq:eff})].
Therefore, $M$ satisfies a quadratic equation whose solution is
\begin{eqnarray}
M&=&\Sigma_1(M)-m_0A_1(M^2)
\nonumber\\
&&{}+\sqrt{m_0^2[1+A_1(M^2)]^2+\Sigma_2(M)[\Sigma_2(M)
-2m_0A_2(M^2)]}.
\label{eq:sol}
\end{eqnarray}
In Eq.~(\ref{eq:sol}) we have chosen the positive square root to ensure that
in the parity-conserving case, where $\Sigma_2(p)=A_2(p^2)=0$,
Eq.~(\ref{eq:sol}) reduces to $M=m_0+\Sigma_1(M)$, which is the correct
expression, as $\Sigma_\mathrm{eff}(p)\to\Sigma_1(p)$ in that limit [cf.\
Eq.~(\ref{eq:eff})].
It is easy to verify that Eq.~(\ref{eq:sol}) is equivalent to the alternative
expression
\begin{equation}
M=m_0\sqrt{f(M^2)},
\label{eq:malt}
\end{equation}
which is the zero of $D(p^2)$ in Eq.~(\ref{eq:d}).

Since $\Sigma_2(p)$ and $A_2(p^2)$ are parity-nonconserving amplitudes, they
are of $\mathcal{O}(g^2)$, where $g$ is a generic weak-interaction gauge
coupling.
If terms of $\mathcal{O}(g^8)$ are neglected, Eq.~(\ref{eq:sol}) simplifies to
\begin{equation}
M=m_0+\Sigma_1(M)+\frac{\Sigma_2(M)[\Sigma_2(M)-2m_0A_2(M^2)]}
{2m_0[1+A_1(M^2)]}
+\mathcal{O}(g^8).
\end{equation}
Thus, we see that the parity-nonconserving interactions introduce an
explicit correction of $\mathcal{O}(g^4)$ in $M$.
Of course, there are also corrections of $\mathcal{O}(g^2)$ and higher in
$\Sigma_1(M)$, as well as QCD corrections in the quark cases.

Parameterizing
\begin{equation}
M=m-i\frac{\Gamma}{2},
\label{eq:par}
\end{equation}
and taking the real and imaginary parts of Eq.~(\ref{eq:M}), we obtain
\begin{eqnarray}
m&=&m_0+\re\Sigma_\mathrm{eff}(M),
\label{eq:polemf}\\
\frac{\Gamma}{2}&=&-\im\Sigma_\mathrm{eff}(M),
\label{eq:polewf}
\end{eqnarray}
which are the counterparts of Eqs.~(\ref{eq:polem}) and (\ref{eq:polew}) and
define the pole mass $m$ and the pole width $\Gamma$ of the unstable fermion.
In analogy with Eqs.~(\ref{eq:mwz}) and (\ref{eq:z}), Eq.~(\ref{eq:polewf})
can be rewritten as
\begin{eqnarray}
\frac{\Gamma}{2}&=&-Z\im\Sigma_\mathrm{eff}(m),
\label{eq:wz}\\
Z&=&\frac{1}{1+\im[\Sigma_\mathrm{eff}(M)-\Sigma_\mathrm{eff}(m)]/(\Gamma/2)}.
\label{eq:zf}
\end{eqnarray}
Indeed, inserting Eq.~(\ref{eq:polewf}) into Eq.~(\ref{eq:zf}),
Eq.~(\ref{eq:wz}) becomes a mathematical identity.

Returning to Eq.~(\ref{eq:s}) and using $a_++a_-=1$ and $a_+-a_-=\gamma_5$,
we write the propagator in the form
\begin{eqnarray}
iS(p)&=&i[S_+(p)a_++S_-(p)a_-],
\nonumber\\
S_\pm(p)&=&\frac{1\mp\Sigma_P(p)}{\notp-m_0-\Sigma_\mathrm{eff}(p)}.
\end{eqnarray}
Dividing numerator and denominator by $1\mp\Sigma_P(p)$, we note that
$S_\pm(p)$ can be expressed as
\begin{eqnarray}
S_\pm(p)&=&\frac{1}{\notp-m_0-\Sigma_\pm^\mathrm{eff}(p)},
\label{eq:spm}\\
\Sigma_\pm^\mathrm{eff}(p)&=&\Sigma_\mathrm{eff}(p)\mp
\frac{[\notp-m_0-\Sigma_\mathrm{eff}(p)]\Sigma_P(p)}{1\mp\Sigma_P(p)}.
\end{eqnarray}
Thus, in the denominator of Eq.~(\ref{eq:spm}) the effective self-energy
$\Sigma_\mathrm{eff}(p)$ has been replaced by $\Sigma_\pm^\mathrm{eff}(p)$.
We note, however, that the two self-energies coincide at $\notp=M$, on account
of Eq.~(\ref{eq:M}).
Namely, we have
\begin{equation}
\Sigma_\pm^\mathrm{eff}(M)=\Sigma_\mathrm{eff}(M).
\label{eq:seffpm}
\end{equation}

In order to construct the renormalized propagator,
\begin{equation}
iS^{(r)}(p)=i\left[S_+^{(r)}(p)a_++S_-^{(r)}(p)a_-\right],
\label{eq:sr}
\end{equation}
it is convenient to use the representation of the unrenormalized amplitude
$S(p)$ given in Eq.~(\ref{eq:inv}).
Recalling that $S(p)$ is the Fourier transform of
$\langle0|T[\psi(x)\overline{\psi}(0)]|0\rangle$ and splitting
$\psi(x)=\psi_+(x)+\psi_-(x)$ into right- and left-handed components
$\psi_\pm(x)=a_\pm\psi(x)$, one finds that the contributions of
$\langle0|T[\psi_-(x)\overline{\psi_-}(0)]|0\rangle$,
$\langle0|T[\psi_+(x)\overline{\psi_-}(0)]|0\rangle$,
$\langle0|T[\psi_+(x)\overline{\psi_+}(0)]|0\rangle$, and
$\langle0|T[\psi_-(x)\overline{\psi_+}(0)]|0\rangle$ are given by the 1st,
2nd, 3rd, and 4th terms of Eq.~(\ref{eq:inv}), respectively.
Shifting $\psi_\pm(x)$ as
\begin{equation}
\psi_\pm(x)=\sqrt{Z_\pm}\psi_\pm^\prime(x),
\end{equation}
where $\psi_\pm^\prime(x)$ are the renormalized fields, it follows that the
renormalized expressions are obtained by dividing the 1st, 2nd, 3rd, and 4th
terms of Eq.~(\ref{eq:inv}) by $|Z_-|$, $\sqrt{Z_+Z_-^*}$, $|Z_+|$, and
$\sqrt{Z_-Z_+^*}$, respectively.
Thus, we obtain
\begin{equation}
S_\pm^{(r)}(p)=\frac{\notp[1-B_\pm(p^2)]+\sqrt{Z_\mp/Z_\pm}m_0[1+A_\mp(p^2)]}
{|Z_\mp|D(p^2)}.
\label{eq:spmr}
\end{equation}
Evaluating the inverse of Eq.~(\ref{eq:sr}), $S^{(r)}(p)$ can also be written
as
\begin{eqnarray}
S^{(r)}(p)&=&\frac{1}{I_+^{(r)}(p)a_++I_-^{(r)}(p)a_-},
\nonumber\\
I_\pm^{(r)}(p)&=&|Z_\pm|\!\left\{\!\notp[1-B_\pm(p^2)]
\!-\!\sqrt{\frac{Z_\mp^*}{Z_\pm^*}}m_0[1+A_\pm(p^2)]\!\right\}\!.
\nonumber\\
&&\label{eq:spmrinv}
\end{eqnarray}

In the case of stable fermions, it is customary to define the
field-renormalization constants so that the pole residue in $S^{(r)}(p)$
equals unity.
In the case of unstable fermions, this is generally not possible, since the
analysis involves two complex functions $S_\pm^{(r)}(p)$ that would require
four constants, while the adjustable constants at our disposal, $Z_-/Z_+$,
$|Z_+|$, and $|Z_-|$, allow for only three independent real parameters.
A particularly simple example of this restriction is provided by unstable
fermions in parity-conserving theories. In that case, there is a single
self-energy $\Sigma(p)$, the residue of the pole is $1/[1-\Sigma^\prime(M)]$,
a complex amplitude, while the field-renormalization constant $Z$ is real.
Returning to the parity-nonconserving theories, and taking these observations
into account, we normalize $S_\pm^{(r)}(p)$ by requiring that the absolute
values of their pole residues equal unity.
A simple and symmetric determination of $Z_\pm$ that satisfies these
constraints is
\begin{eqnarray}
Z_\pm&=&\frac{1+R_\pm}{2F[1-B_\pm(M^2)]},
\nonumber\\
R_\pm&=&\frac{1+A_\pm(M^2)}{1+A_\mp(M^2)},
\nonumber\\
F&=&1-M^2\frac{f^\prime(M^2)}{f(M^2)}.
\label{eq:zpm}
\end{eqnarray}
Writing $Z_\pm=|Z_\pm|\mathrm{e}^{i\theta_\pm}$, the residues of
$S_\pm^{(r)}(p)$ are $\mathrm{e}^{i\theta_\mp}$, respectively, in agreement
with our requirements.
This implies that in the resonance region the renormalized propagator behaves
as $i(\mathrm{e}^{i\theta_-}a_++\mathrm{e}^{i\theta_+}a_-)/(\notp-M)$, which,
in leading order, reduces to the Breit-Wigner form $i/(\notp-m+i\Gamma/2)$.

Inserting Eqs.~(\ref{eq:malt}) and (\ref{eq:zpm}) in Eq.~(\ref{eq:spmr}),
$S_\pm^{(r)}(p)$ can be expressed completely in terms of the functions
$A_\pm(p^2)$ and $B_\pm(p^2)$, as
\begin{eqnarray}
S_\pm^{(r)}(p)&=&\frac{2F\mathrm{e}^{i\theta_\mp}}{1+R_\mp}\,
\frac{1-B_\mp(M^2)}{1-B_\mp(p^2)}
\nonumber\\
&&{}\times
\frac{\notp+M[1+A_\mp(p^2)][1-B_\pm(M^2)]/\{[1+A_\pm(M^2)][1-B_\pm(p^2)]\}}
{p^2-M^2f(p^2)/f(M^2)}.
\label{eq:spmrfin}
\end{eqnarray}
In the case of CP conservation, we have $A_+(p^2)=A_-(p^2)$ and $R_\pm=1$, so
that Eqs.~(\ref{eq:zpm}) and (\ref{eq:spmrfin}) simplify considerably.

In summary, in the approximation of neglecting flavor mixing, we have derived
general and closed expressions for the pole mass and width of unstable
fermions in parity-nonconserving gauge theories [Eqs.~(\ref{eq:polemf}) and
(\ref{eq:polewf})], their unrenormalized and renormalized propagators
[Eqs.~(\ref{eq:pro}), (\ref{eq:inv}), (\ref{eq:s}) and
Eqs.~(\ref{eq:spmr}), (\ref{eq:spmrinv}), (\ref{eq:spmrfin})], and their
field-renormalization constants [Eq.~(\ref{eq:zpm})]. 
We also note that the discussion after Eq.~(\ref{eq:zpm}) provides a
theoretical framework to employ $i/(\notp-m+i\Gamma/2)$ as leading-order
propagator, particularly in the resonance region.
In turn, it was emphasized in Ref.~\cite{Passera:1998uj} that the systematic
use of this propagator in the evaluation of the gluonic and photonic
contributions to the fermion self-energy avoids the emergence in the resonance
region of catastrophic on-shell singularities proportional to
$[m_\mathrm{os}\Gamma_\mathrm{os}/(p^2-m_\mathrm{os}^2)]^n$ $(n=2,3,\ldots)$.
Furthermore, in the quark case, the same diagrams lead to an unbounded gauge
dependence of $\mathcal{O}(\alpha_s(m)\Gamma)$ in the on-shell mass
$m_\mathrm{os}$, which is neatly avoided by employing the pole mass $m$.

The significance of the concepts of pole mass and width we have discussed is
that they are gauge independent, and thus satisfy a fundamental tenet of gauge
theories to be identified with physical observables.

The work of B.A.K. was supported in part by the German Research Foundation
through the Collaborative Research Center No.\ 676 {\it Particles, Strings and
the Early Universe---the Structure of Matter and Space-Time}.
The work of A.S. was supported in part by the National Science Foundation
Grant No.\ PHY-0245068.

\end{document}